\documentclass{aa}  

\usepackage{graphicx}
\usepackage{amsmath}	
\usepackage{amssymb}	
\usepackage{txfonts}
\usepackage{natbib}
\usepackage{longtable}
\usepackage[normalem]{ulem}

\newcommand{\teff}{${T}_{\mathrm{eff}}$}
\newcommand{\logg}{$\log{g}$}
\newcommand{\msun}{$M_{\odot}$}
\newcommand{\rsun}{$R_{\odot}$}

\begin{document} 
	
	\title{The McDonald Observatory search for pulsating sdA stars: asteroseismic support for multiple populations}
	
	\author{Keaton~J.~Bell,\inst{1,2}
		Ingrid~Pelisoli,\inst{3}
		S.~O.~Kepler,\inst{3}
		W.~R.~Brown,\inst{4}
		D.~E.~Winget,\inst{5,6}
		K.~I.~Winget,\inst{5,6}
		Z.~Vanderbosch,\inst{5,6}
		B.~G.~Castanheira,\inst{5,6,7}
		J.~J.~Hermes,\inst{8,9}
		M.~H.~Montgomery\inst{5,6}
		and D.~Koester\inst{10}
	}
	
	\institute{Max-Planck-Institut f{\"u}r Sonnensystemforschung (MPS), Justus-von-Liebig-Weg 3, 
		37077 G{\"o}ttingen, Germany\\
		\email{bell@mps.mpg.de}
		\and
		Department of Physics and Astronomy, Stellar Astrophysics Centre, Aarhus University, Ny Munkegade 120, 8000 Aarhus C, Denmark
		\and
		Instituto de F{\'i}sica, Universidade Federal do Rio Grande do Sul, 91501-970 Porto-Alegre, RS, Brazil             
		\and
		Smithsonian Astrophysical Observatory, Cambridge, MA\,-\,02138, USA
		\and
		Department of Astronomy, University of Texas at Austin, Austin, TX\,-\,78712, USA
		\and
		McDonald Observatory, Fort Davis, TX\,-\,79734, USA
		\and
		Baylor University, Waco, TX\,-\,76798, USA
		\and
		Department of Physics and Astronomy, University of North Carolina, Chapel Hill, NC\,-\, 27599, USA
		\and
		Hubble Fellow
		\and
		Institut für Theoretische Physik und Astrophysik, Universität Kiel, D-24098 Kiel, Germany
	}
	
	\date{}

	\titlerunning{Pulsation timescales of sdA stars}
	\authorrunning{Bell et al.}
	
	
	\abstract
	{The nature of the recently identified ``sdA'' spectroscopic class of star is not well understood. The thousands of known sdAs have H-dominated spectra, spectroscopic surface gravities intermediate to main sequence stars and isolated white dwarfs, and effective temperatures below the lower limit for He-burning subdwarfs.  Most are likely products of binary stellar evolution, whether extremely low-mass white dwarfs and their precursors, or blue stragglers in the halo.}
	{Stellar eigenfrequencies revealed through time series photometry of pulsating stars sensitively probe stellar structural properties.  The properties of pulsations exhibited by any sdA stars would contribute importantly to our developing understanding of this class.}
	{We extend our photometric campaign to discover pulsating extremely low-mass white dwarfs from McDonald Observatory to target sdA stars classified from SDSS spectra. We also obtain follow-up time series spectroscopy to search for binary signatures from four new pulsators.}
	{Out of 23 sdA stars observed, we clearly detect stellar pulsations in seven. Dominant pulsation periods range from 4.6 minutes to 12.3 hours, with most on $\sim$hour timescales. We argue specific classifications for some of the new variables, identifying both compact and likely main sequence dwarf pulsators, along with a candidate low-mass RR Lyrae star.}
	{With dominant pulsation periods spanning orders of magnitude, the pulsational evidence supports the emerging narrative that the sdA class consists of multiple stellar populations. Since multiple types of sdA exhibit stellar pulsations, follow-up asteroseismic analysis can be used to probe the precise evolutionary natures and stellar structures of these individual subpopulations.}
	
	
	\maketitle
	%
	
	\section{Introduction}
	
	Model atmosphere fits to spectra from Sloan Digital Sky Survey (SDSS) data releases (DR) 10 and 12 revealed a population of thousands of stars with H-dominated spectra, effective temperatures $T_\mathrm{eff}\leq 20{,}000$\,K (majority with 9000\,K $> T_\mathrm{eff} >$ 6500\,K), and surface gravities $6.5 > \log{g} > 5.5$ \citep{Kepler2015,Kepler2016}. This gravity range is greater than for main sequence stars, but less than for white dwarfs (WDs) that formed through single-star evolution given the age of the Galaxy \citep{Kilic2007}. Hot sdB and sdO sub\-dwarf stars are often observed with these gravities, but they must have $T_\mathrm{eff}\gtrsim20{,}000$ to fuse He in their cores above the zero-age horizontal branch \citep[for a recent review of subdwarfs, see][]{Heber2016}. Extending the classification scheme for subdwarf stars, \citet{Kepler2016} applied the label ``sdA'' to this newly uncovered stellar population that shares an effective temperature range with main-sequence A stars. This classification is based solely on spectroscopically determined parameters and does not require an evolutionary origin similar to the hot, He-burning subdwarfs.
	
	Another class of star is known to populate this \logg--\teff\ regime: extremely low-mass (ELM) WDs. The exact definition of an ELM WD varies in the literature, but they are generally understood to have degenerate He cores that formed as the result of mass loss in tight binaries. Interest in these objects has been spurred by recent observational discovery and characterization, especially by the dedicated ELM Survey \citep[][and references therein]{Brown2016}.  Out of 78 objects cataloged by the ELM Survey to have spectroscopic H-line profiles that agree with ELM WD atmosphere models, radial velocity variations confirm that 67 are in close binary systems \citep{Brown2016}, supporting the binary mass-transfer formation theory.  
	
	Despite the overlap in spectroscopic properties of the sdA and ELM WDs, the population size and other observational properties of sdA stars are not consistent with the hypothesis that they are simply older, cooler ELM WDs. \citet{Hermes2017} appealed primarily to radial velocity variations of SDSS subspectra to argue that the vast majority of sdA stars are not in short-period binary systems, and therefore are not created through the same evolutionary pathways as the ELM Survey objects. \citet{Brown2017} consider model ELM WD evolutionary sequences \citep{Althaus2013,Istrate2016a} to support this argument, showing that the concentration of sdA stars below 9000\,K is too large to be consistent with further evolved ELM WDs. Comparing survey colors and reduced proper motions of the two samples, \citet{Hermes2017} and \citet{Brown2017} both argue that most sdAs are main-sequence A and F stars. Specifically, \citet{Brown2017} demonstrate that fitting main sequence model spectra with pure-H atmosphere models can cause \logg\ values to be systematically overestimated by up to 1\,dex. By constraining the radii of six sdAs in eclipsing binaries with time series spectroscopy and Catalina Sky Survey \citep{Drake2009} photometry, \citet{Brown2017} definitively show that at least a few specific sdA stars are too large to be cooling-track ELM WDs. 
	
	\citet{Pelisoli2018a} recently provided a thorough treatment of the problem of placing the sdA stars into a specific evolutionary context. They present the arguments for and against each of four physical interpretations:
	\begin{itemize}
		\item \textbf{ELM WDs or their precursors (pre-ELM WDs):} after binary interactions strip their atmospheres, He-core star models contract and then cool through the spectroscopic parameter space of sdA stars, with CNO-burning flashes possibly causing some to loop repeatedly through this space before settling onto final cooling tracks \citep[e.g.,][]{Althaus2013,Istrate2016a}. We noted above how the properties and large number of sdAs do not agree with the hotter sample of ELM Survey objects \citep{Brown2016}.
		\item \textbf{Blue stragglers in the halo:} \citet{Brown2017} proposed that sdAs are mostly main sequence A stars with systematically overestimated \logg\ values. Assuming main sequence radii, the spatial velocities of sdAs would place most of them in the Galactic halo, although the velocities of 35\% would be too high even for this hypothesis \citep{Pelisoli2018a}. Since the main sequence lifetime of an A star is an order of magnitude shorter than the age of the halo population, these would have to be rejuvenated blue straggler stars. 
		\item \textbf{Late-type main sequence stars:} dwarf stars of type F and later with $T_\mathrm{eff} \lesssim 7500$\,K have main sequence lifetimes that are long enough to be consistent with the halo age. This could explain the cooler concentration of sdA stars that is revealed through density estimation of ($g-z$) colors \citep{Pelisoli2018a}, but this supposes that spectroscopic \logg\ values of dwarf stars are being systematically overestimated.
		\item \textbf{Main sequence stars with hot subdwarf companions:} source blending in main sequence + subdwarf binaries could yield spectroscopic \logg\ measurements in the observed range.  The main sequence star would dominate the optical flux, but the subdwarf would produce a UV excess, which \citet{Pelisoli2018a} find for only 0.5\% of sdAs with GALEX photometry.
	\end{itemize}
	While a consensus on the exact nature of sdA stars remains out of reach, the full range of possibilities is gradually coming into focus. The main conclusion of \citet{Pelisoli2018a} is that the sdAs must consist of multiple populations in order to explain all of the observations. \citet{Pelisoli2018c} used parallaxes from \emph{Gaia} DR2 to argue that hundreds of sdAs are too small to be main sequence dwarfs, but must instead be ELM WDs or their precursors. Whatever the formation channels, sdA stars represent interesting stellar populations, and better constraints on their properties will guide our theories of stellar evolution.

	\begin{table*}
		\caption{Best-fit spectroscopic parameters of observed sdA stars}              
		\label{tab:targets}      
		\centering                                      
		\begin{tabular}{l l l l l l l}  
			\hline\hline
			SDSS object name & Plate-MJD-Fiber & SDSS $g$ & \teff & \logg & \teff & \logg\\
			&  & (mag) & (K; solar z)\tablefootmark{a} & (cgs; solar z)\tablefootmark{a} & (K; pure H)\tablefootmark{a} & (cgs; pure H)\tablefootmark{a}\\
			\hline
			J123831.40$-$014654.3 &0335-52000-0497 & 16.97 &7932(17) &4.836(0.081)& 7864&  5.03\\
			J131011.61$-$014233.0 &0340-51990-0426 & 16.53 &8237(10) &4.721(0.060)& 8224&  5.33\\
			J221524.54$-$005018.2 &0374-51791-0131 & 17.97 &7763(90)\tablefootmark{b} &4.529(0.038)\tablefootmark{b}& 7838&  5.72\\
			J090804.54$-$000208.8\tablefootmark{c} &0470-51929-0076 & 15.13 &8168(10) &5.039(0.036)& 8130&  5.33\\
			J091416.42+004146.8  &0472-51955-0499 & 17.39 &8417(20) &5.883(0.081)& 8440&  5.71\\
			J093003.42+054815.5  &0992-52644-0435 & 18.13 &8760(37) &6.240(0.105)& 8656&  5.52\\
			J074939.74+194203.5  &1582-52939-0406 & 18.04 &7846(32) &5.958(0.094)&  8042  &  6.71\\
			J162953.16+220634.5  &1658-53240-0584 & 18.20 &8087(35) &5.833(0.122)& 8057&  5.08\\
			J160410.81+062705.5 &1729-53858-0213 & 15.10 &7969(9) &5.473(0.039)& 8097&  5.71\\
			J082900.96+084645.4  &1758-53084-0449 & 17.81 &8394(25) &5.995(0.100)& 8195&  5.16\\
			J140119.77+351323.2 &1838-53467-0506 & 17.64 &8054(19) &4.996(0.094)& 7908&  5.21\\
			J192253.84+783959.1 &1857-53182-0494 & 14.32 &7980(10) &5.328(0.032)& 7904&  5.24\\
			J233258.96+490400.3 &1888-53239-0477 & 16.26 &8323(12) &6.210(0.036)& 8283&  5.77\\
			J223831.92+125318.3 &1892-53238-0249 & 15.54 &7980(11) &4.816(0.056)& 7999&  5.03\\
			J114224.61+374703.7 &1997-53442-0225 & 15.09 &8075(9) &5.211(0.037)& 8082&  5.16\\
			J093356.45+191601.5  &2361-53762-0308 & 18.54 &8333(39) &6.088(0.127)& 8126&  3.75\\
			J083054.47$-$035118.9 &2807-54433-0499 & 15.15 &8473(8) &4.444(0.042)& 8460&  5.03\\
			J113143.46$-$074220.5 &2861-54583-0078 & 15.93 &7853(12) &4.635(0.064)& 7637&  4.61\\
			J045309.80$-$041800.7 &3123-54741-0483 & 16.99 &7897(7) &4.693(0.042)& 7665&  4.59\\
			J074013.22+481036.7 &3668-55478-0370  & 16.68 &  7961(5) & 3.932(0.021)  & 7748 & 5.34\\
			J075644.33+502741.2 &3679-55209-0229   & 16.56 & 7108(7) & 4.045(0.027)  & 6875 & 5.34\\
			J223716.61+052228.3 &4291-55525-0872   & 19.46 & 8902(40) & 6.695(0.017)  & 8856 &6.66\\
			J161831.69+385415.2 &5189-56074-0177  & 19.67 &  9307(70) & 7.153(0.096)  & 9149 &6.70\\
			J161831.69+385415.2\tablefootmark{d} & 5199-56067-0744  & 19.67 & 9354(54) & 6.257(0.075)  & 9058  & 6.04 \\
			\hline
		\end{tabular}
		\tablefoot{
			\tablefoottext{a}{We quote internal statistical uncertainties. The \teff\ and \logg\ uncertainties for the pure-H models are similar to the solar z uncertainties. An analysis by \citet{Kepler2016} of the systematic uncertainties from multiple spectra indicated 5\% in \teff\ and 0.05\,dex in \logg.}
			\tablefoottext{b}{The fitting pipeline of \citet{Kepler2016} does not provide uncertainties for this fit because it does not agree with the cool \teff\ implied by the SDSS photometry for this star. We estimate the uncertainties by taking the difference between the parameters of the best fits to the tabulated Plate-MJD-Fiber and a second SDSS spectrum of this object, 3146-54773-0003.}
			\tablefoottext{c}{Eclipsing binary with 7.68-hour orbit (see text).}
			\tablefoottext{d}{This is a second SDSS spectrum of the same object as above.}
		}
	\end{table*}
	
	\subsection{Pulsations in the sdA regime}
	
	The potential for ELM WDs to improve our understanding of binary, mass-transfer evolution was heightened by the discovery that many exhibit photometric variations as they pulsate in a low-\logg\ extension of the ZZ Ceti (H-atmosphere WD) instability strip.  Analysis of the pulsational eigenfrequencies can help to constrain the details of their interior stellar structures.  We discovered the first five pulsating ELM WDs with the McDonald Observatory 2.1-meter Otto Struve Telescope \citep{Hermes2012,Hermes2013a,Hermes2013b} from candidates identified in the ELM Survey.  The discovery of the first pulsating ELM WD to be identified from SDSS spectra was presented in conference proceedings by \citet[][]{Bell2015}, and we include this object in the present paper. \citet{Kilic2015} published a seventh pulsating ELM WD that is the binary companion of a millisecond pulsar.  
	
	\citet{Bell2017} discovered three additional pulsating stars from the two most recent ELM Survey publications \citep{Gianninas2015,Brown2016}.  All three are in the region of overlap between the ELM Survey catalog and the sdAs found in SDSS spectra to have $T_\mathrm{eff}\lesssim9000$\,K, and none show short-period radial velocity variations as required by some more conservative definitions of an ELM WD. \citet{Bell2017} argue that these pulsating stars are consistent with belonging to the same population as the sdAs.  One of the targets exhibits a dominant pulsation period of 4.3 hours, which exceeds the upper limit for surface reflection in a cooling-track WD \citep{Hansen1985}.
	
	\citet{Maxted2013,Maxted2014}, \citet{Corti2016}, \citet{Zhang2016}, and \citet{Gianninas2016} have also reported pulsations in pre-ELM stars that evolve through the region of sdAs as they contract toward their final cooling tracks. Some of these objects pulsate at higher temperatures (12{,}000\,K $\gtrsim T_\mathrm{eff} \gtrsim11{,}000$\,K) than a simple extrapolation of the ZZ Ceti instability strip to low \logg, which may be due to the contribution of atmospheric helium to the driving \citep{Jeffery2013,Corsico2016b,Gianninas2016,Fontaine2017}. However, the claimed pulsating pre-ELMs from \citet{Corti2016} are sdA stars from SDSS spectra, and the authors suggest that the stars may be high amplitude delta Scuti pulsators instead. While \citet{Zhang2016} claim to have observed gravity-modes in the \emph{Kepler} data of a pre-ELM WD in an eclipsing binary, the relationship they find between the rotational splitting of pulsation modes and the stellar rotation rate of the star is characteristic of pressure modes (p-modes). We suspect that this system is more similar to the pre-ELM WD plus pulsating delta Scuti binaries of \citet{Zhang2017}.
	
	The many observational discoveries and measurements of pulsating stars in the parameter space of ELM and pre-ELM WDs has prompted extensive theoretical investigations into their driving mechanisms and asteroseismic interpretations \citep{Corsico2012,Corsico2016b,Corsico2014a,Corsico2014b,Corsico2016a,Istrate2016,Istrate2017,Fontaine2017,Calcaferro2017a,Calcaferro2017b,SA2018}.
	
	Given the overlap in parameter space of the sdAs with the ELM WD instability strip, plus the evidence from \citet{Bell2017} that even non-cooling-track-WDs pulsate in this domain, we extended our observational search for pulsating stars at McDonald Observatory to include sdA targets.  The overall timescales of sdA pulsations would give a first-order indication of the types of stars that exist in this population.  Detailed follow-up asteroseismic interpretation of specific pulsational eigenfrequencies will go further to constrain the interior structures of these enigmatic stars.
	
	We describe the photometric observations of 23 sdA stars from McDonald Observatory, as well as the time series spectroscopic follow-up on four in Section~\ref{sec:obs}.  In Section~\ref{sec:anal}, we present the frequency analysis of seven sdA stars that clearly exhibit pulsations in the light curves, as well as the orbital constraints from spectroscopy, studying the natures of individual objects.  We place these stars in the context of other pulsating variables in the concluding Section~\ref{sec:disc}.  We argue that the large range of dominant pulsation periods observed, from 4.6 minutes to 12.3 hours, supports that the sdAs are made up of multiple stellar populations that pulsate in this region of \logg--\teff\ space.

	\section{Observations}
	\label{sec:obs}
	
	\subsection{Time series photometry}
	We obtained 196 hours of time series photometry on 23 stars near the low-\logg\ extension of the ZZ Ceti instability strip.  Unlike the systematic search for photometric variability of ELM Survey objects described in \citet{Bell2017}, it was not practical to carry out such an exhaustive follow-up study of the thousands of sdA stars classified from SDSS spectroscopy. Our target selection was notably nonuniform, and it was often informed by parameters from preliminary fits to the SDSS spectra as we were beginning to make sense of the data.  Table~\ref{tab:targets} provides the best-fit spectroscopic parameters from the two sets of models described in \citet{Pelisoli2018a}: pure hydrogen atmospheres (pure H), and models matching the composition of the solar atmosphere (solar z).

	\begin{table}
		\caption{Journal of photometric observations}
		\label{tab:observinglog}
		\centering
		\begin{tabular}{l c c c c r}
			\\
			\hline
			SDSS & Date & Exposure & Run Duration \\  
			& (UTC) & Time (s)	& (h) \\
			\hline
			
			J1618+3854    & 23 Apr 2014 & 20 & 6.5 \\*
			& 24 Apr 2014 & 25 & 6.4 \\*
			& 22 May 2014 & 30 & 1.0 \\*
			& 22 May 2014 & 20 & 3.9 \\*
			& 24 May 2014 & 30 & 2.6 \\*
			& 25 May 2014 & 30 & 5.9 \\*
			& 27 May 2014 & 30 & 7.8 \\*
			& 21 Apr 2015 & 20 & 4.6 \\*
			& 26 Apr 2015 & 30 & 5.7 \\*
			& 20 May 2015 & 20 & 3.2 \\*
			& 26 May 2015 & 30 & 7.7 \\*
			J1131$-$0742     & 19 Jan 2015 & 10 & 3.7 \\*
			& 20 Jan 2015 & 5 & 5.0 \\*
			J0756+5027    & 11 Feb 2015 & 1 & 7.1 \\*
			J1310$-$0142  & 14 Mar 2015 & 10 & 3.4 \\*
			& 21 Jun 2017 & 15 & 1.5 \\*
			& 21 Jun 2017 & 30 & 0.8 \\*
			& 22 Jun 2017 & 15 & 2.6 \\*
			& 23 Jun 2017 & 30 & 1.3 \\*
			J1142+3747    & 25 Apr 2015 & 3 & 2.0 \\*
			& 22 May 2017 & 15 & 3.2\\*
			& 24 May 2017 & 30 & 3.2\\*
			J2238+1253    & 11 Aug 2015 & 5 & 4.5  \\*
			& 12 Aug 2015 & 5 & 1.6 \\*
			& 13 Aug 2015 & 5 & 0.8 \\*
			& 14 Aug 2015 & 3 & 5.1 \\*
			& 16 Aug 2015 & 3 & 1.0 \\*
			J1604+0627    & 14 Sep 2015 & 3 & 1.7 \\*
			& 16 Sep 2015 & 5 & 2.3 \\*
			& 06 Jul 2016 & 1 & 3.1 \\*
			\hline
			J2237+0522    & 25 Jun 2014 & 30 & 3.2 \\*
			& 26 Jun 2014 & 30 & 3.0 \\*
			& 12 Sep 2015 & 60 & 5.0 \\*
			& 13 Sep 2015 & 20 & 6.1 \\*
			J0453$-$0418     & 24 Jan 2015 & 10 & 2.7 \\*
			J0830$-$0351     & 24 Jan 2015 & 10 & 2.5 \\*
			J0740+4810    & 26 Jan 2015 & 15 & 3.6 \\*
			J1238$-$0146     & 15 Mar 2015 & 30 & 3.3 \\*
			J0908$-$0002     & 26 Apr 2015 & 3 & 2.8 \\*
			J2215$-$0050     & 14 Sep 2015 & 20 & 5.2 \\*
			& 15 Sep 2015 & 5 & 6.3 \\*
			J2332+4904    & 16 Sep 2015 & 5 & 5.6 \\*
			J1922+7839    & 17 Sep 2015 & 3 & 3.9 \\*
			J1401+3513    & 07 Jan 2016 & 5 & 2.5 \\*
			J1629+2206    & 06 Aug 2016 & 5 & 4.0 \\*
			& 07 Aug 2016 & 5 & 4.7 \\*
			J0749+1942    & 28 Nov 2016 & 10 & 1.5 \\*
			& 28 Nov 2016 & 30 & 2.5 \\*
			& 29 Nov 2016 & 20 & 3.1 \\*
			J0829+0846    & 01 Jan 2017 & 10 & 2.0 \\*
			& 02 Jan 2017 & 10 & 4.8 \\*
			J0914+0041    & 20 Jan 2017 & 5 & 1.3 \\*
			& 20 Jan 2017 & 5 & 1.1 \\*
			J0933+1916    & 26 Jan 2017 & 20  & 3.7 \\*
			J0930+0548    & 03 Mar 2017 & 14  & 3.0\\
			\hline
			
		\end{tabular}
	\end{table}

	The details of our observing runs are logged in Table~\ref{tab:observinglog}. Our observational setup and data reduction pipeline was as described in \citet{Bell2017}. We observed our targets with the frame-transfer ProEM CCD camera on the McDonald 2.1-meter Otto Struve Telescope through a BG40 filter to reduce sky noise. All frames were dark-subtracted and flatfielded with nightly calibration frames using standard {\sc IRAF} tasks.  We measure circular-aperture photometry to obtain light curves for the target and bright field stars with the {\sc IRAF} script {\sc ccd\_hsp} \citep{Kanaan2002}.  Then we divided each target light curve by the weighted mean comparison star light curve, clipped extreme outliers, divided low-order polynomials to remove long-timescale airmass effects, and applied a barycentric correction to the timestamps with the {\sc WQED} software \citep{wqed}.
	
	\subsection{Time domain spectroscopy}
	
	The discovery of photometric variability prompted us to obtain follow-up spectroscopy of four objects, using the same instrumentation, reduction, and analysis tools as the ELM Survey \citep{Brown2016}: wavelength calibration against a comparison lamp, flux calibration against blue spectrophotometric standards \citep{Massey1988}, and measurement of radial velocities (RVs) via cross-correlation with spectral templates \citep{Kurtz1998}.
	
	We obtained spectra of SDSS\,J1618+3854 covering 3650--4500\,\AA\ at 1-\AA\ resolution  with the Blue Channel spectrograph \citep{Schmidt1989} on the 6.5-meter MMT.  After two individual observations in July 2016 and March 2017, we obtained ten additional spectra over 25-27 June 2017, and seven more on 13 May 2018.
	
	We obtained 25 spectra of SDSS\,J1131$-$0742 during four nights between 18 February and 18 April 2015 with the FAST instrument \citep{Fabricant1998} on the Fred L.\ Whipple Observatory 1.5-meter telescope.   We also obtained ten back-to-back FAST spectra of SDSS\,J2238+1253 over 2.1 hours on 03 September 2016, and the same for SDSS\,J1142+3747 on 18 June 2017. These observations cover the wavelength range 3600--5400\,\AA\ with 1.7-\AA\ resolution. 
	
	We supplemented the observations of SDSS\,J1131$-$0742 with 2-\AA-resolution spectra covering 3600--4950\,\AA\ from the Goodman Spectrograph \citep{Clemens2004} on the 4.1-meter SOAR telescope. We obtained twelve SOAR spectra on 26 April 2016, and nine more on 31 May 2017. We followed the same basic steps to reduce these data, with a detailed description given in \citet{Pelisoli2018b}.

	\section{Seven New Pulsating sdA Stars}
	\label{sec:anal}
	
	The first seven sdA targets listed in Table~\ref{tab:observinglog} all exhibit clear pulsation signatures in our photometry. We characterize these stars individually in this section, measuring their pulsation periods and searching for RV variations from the stars that we observed spectroscopically. 
	
	All Fourier transforms (FTs) of our light curves were computed with the {\sc Period04} software \citep{Period04} and oversampled by a factor of 20. We follow the same process of signal detection and prewhitening described in \citet{Bell2017}, fitting and subtracting a sum of sinusoids to the most significant signals iteratively until the residuals contain no compelling peaks above a rule-of-thumb $4\langle A\rangle$ significance criterion \citep[where $\langle A\rangle$ is the locally-evaluated mean amplitude of the FT;][]{Breger1993}. 
	
	Gaps in multi-night light curves introduce aliases to our FTs that compete to describe the observed variations.  Very approximately, discrete aliases for observations taken on consecutive nights have a characeristic spacing of $\approx$11.6\,$\mu$Hz and they span a frequency range roughly equal to the inverse of the typical run length, though actual spectral windows can be far more complicated. While the iterative prewhitening procedure can produce a frequency solution that closely matches the data, there are often many viable solutions that employ different aliases.  The highest peaks in the FT do not necessarily coincide with the intrinsic signal frequencies, as \citet{Bell2017b} recently demonstrated in a comparison of ground- and space-based photometry of two pulsating WDs.  We emphasize this point by exploring two candidate solutions for our first sdA pulsator, SDSS\,J1618+3854, below. For the other stars, we only record the frequencies of the highest-amplitude aliases. The uncertainties that we quote are \emph{intrinsic} to the least-squares fitting \citep[given by analytical expressions in][]{Montgomery1999}; they do not account for the additional uncertainty caused by aliasing.
	
	\begin{figure*}
		\centering
		\includegraphics[width=2\columnwidth]{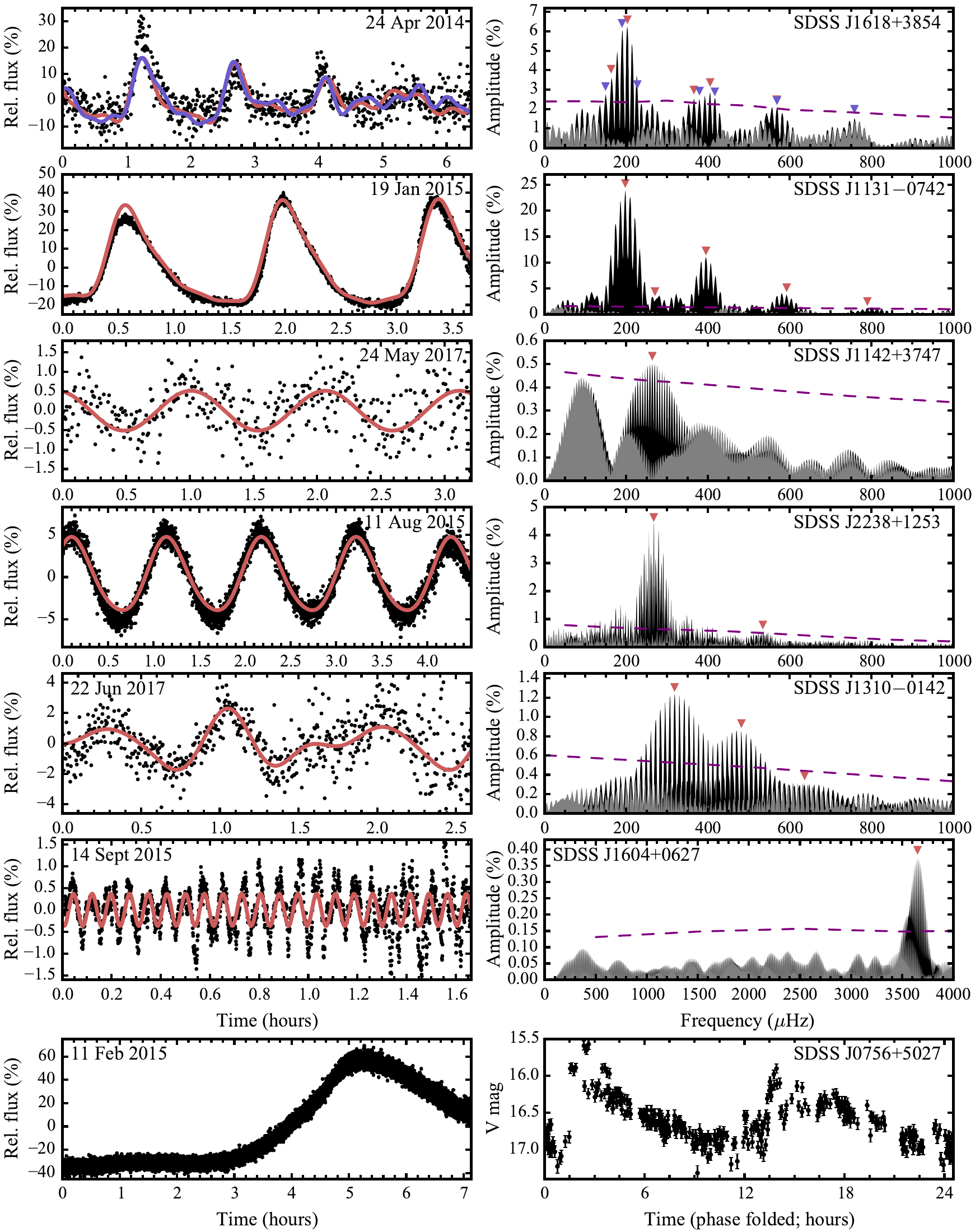}
		\caption{Photometric signatures of pulsations in seven new pulsating sdA stars. The left panels display portions of the McDonald light curves.  The top six right-hand panels display FTs of data subsets (see text).  Frequencies in our adopted solutions are indicated with triangles in the FTs, and the corresponding best fits are plotted over the light curves.  The two frequency solutions for SDSS\,J1618+3854 (see text) are color-coded.   Final significance threshold are marked with dashed lines, and the FTs of the final residuals are displayed in gray. The light curve of SDSS\,J1604+0627 is smoothed over 10 points to guide the eye.  Our McDonald photometry of SDSS\,J0756+5027 does not cover a full pulsation period, so we plot V-band data from the Catalina Sky Survey (see text), folded on twice the dominant period of 12.263\,hours, in the bottom right panel.}
		\label{fig:data}
	\end{figure*}
	
	In our aim to assess the typical pulsation timescales of sdA targets, the tabulated frequencies are perfectly representative.  However, one should ideally account for the full set of possible frequency solutions when comparing against asteroseismic models to constrain the stellar structure. This could be achieved by computing many frequency solutions through a Monte Carlo prewhitening procedure, or by exploring the solution space with Markov Chain Monte Carlo fitting of the light curves directly. To help enable such analyses, we are publishing our reduced light curves for all new pulsators as supplemental online material.
	
 	For the targets that do not exhibit significant variability in our photometry, we only make limited claims about a lack of pulsations. Most are observed on a single night, and destructive interference from beating of closely spaced frequencies could cause a null detection. Variations on a similar timescale to changing airmass or transparency could also mask pulsations.  We indicate approximate limits on the periods and amplitudes of pulsations that may be present for targets that we observed for at least two hours on each of multiple nights in Table~\ref{tab:nov}.  Following \citet[][]{Bell2017}, we assert that each of these stars does not pulsate with periods shorter than the second-shortest run length and with amplitudes greater than the largest $4\langle A\rangle$ measured in any 200-$\mu$Hz-wide window in the FT of any nightly light curve.  We note that we did observe evidence of an eclipse in the light curve of SDSS\,J0908-0002. Photometry from the Asteroid Terrestrial-impact Last Alert System Survey \citep[ATLAS;][]{Heinze2018} and the Catalina Sky Survey \citep[CSS;][]{Drake2009} reveals eclipses and ellipsoidal variations corresponding to a 7.68-hour orbit, lending support to a mass-transfer evolutionary history for this object.
 	
	\begin{table}[t]
		\centering
		\caption{Limits on pulsations in sdA stars}
		\label{tab:nov}
		\begin{tabular}{lccc} 
			\hline
			SDSS & Period & or & Amplitude\\
			& (hr) & & (\%)\\
			\hline
			J2237+0522    & $>5.0$ &  & $<2.5$ \\*
			J2215$-$0050  & $>5.2$ &  & $<0.6$ \\*
			J1629+2206    & $>4.0$ &  & $<1.1$ \\*
			J0749+1942    & $>3.1$ &  & $<1.1$ \\*
			J0829+0846    & $>2.0$ &  & $<1.3$ \\*
			\hline
			
		\end{tabular}
	\end{table}
	
	\subsection{SDSS\,J1618+3854}
	
	\begin{table}[b]
		\centering
		\caption{Two possible frequency solutions for the April 2014 observations of SDSS\,J1618+3854}
		\label{tab:sol1}
		\begin{tabular}{lccr} 
			\hline
			Mode ID  & Frequency & Period & Amplitude\\
			& ($\mu$Hz) & (s) & (\%)\\
			\hline
			$f_1$ & 202.61(15) & 4936(4) & 6.19(19)\\
			$f_2$ & 162.8(3) & 6143(11) & 3.35(19)\\
			$2f_1$ & 405.2(3) & 2468(18) & 2.66(19)\\
			$f_1+f_2$ & 365.4(3) & 2740(20) & 2.68(19)\\
			$2f_1+f_2$ & 568.0(4) & 1761(12) & 1.88(19)\\
			\hline
			$f_1$ & 189.56(16) & 5275(5) & 5.63(19)\\
			$f_2$ & 226.7(3) & 4410(7) & 2.76(19)\\
			$f_3$ & 149.1(4) & 6706(16) & 2.59(19)\\
			$2f_1$ & 379.1(4) & 2638(3) & 2.38(19)\\
			$f_1+f_2$ & 417.2(4) & 2397(2) & 2.21(19)\\
			$3f_1$ & 568.7(6) & 1758.4(17) & 1.69(19)\\
			$4f_1$ & 748.3(7) & 1318.8(12) & 1.37(19)\\
			\hline
		\end{tabular}
	\end{table}
	
	The first pulsating sdA star identified from SDSS DR10 spectroscopy, SDSS\,J1618+3854, was presented as the sixth pulsating ELM WD in conference proceedings by \citet{Bell2015}. The realization from SDSS DR12 that this may belong to the sdA population caused us to rethink this classification.
	
	The combined light curve of the two long discovery runs on 23 and 24 April 2014 remains the easiest to interpret due to its relatively simple spectral window. Still, diurnal aliasing confuses our frequency measurements.  We demonstrate the severity of this effect by presenting two of many viable frequency solutions here. A portion of the April 2014 light curve and its FT are displayed in the first row of Figure~\ref{fig:data}.

	\begin{figure}
		\centering
		\includegraphics[width=0.9\columnwidth,trim={0 6.4cm 1.5cm 10.0cm},clip]{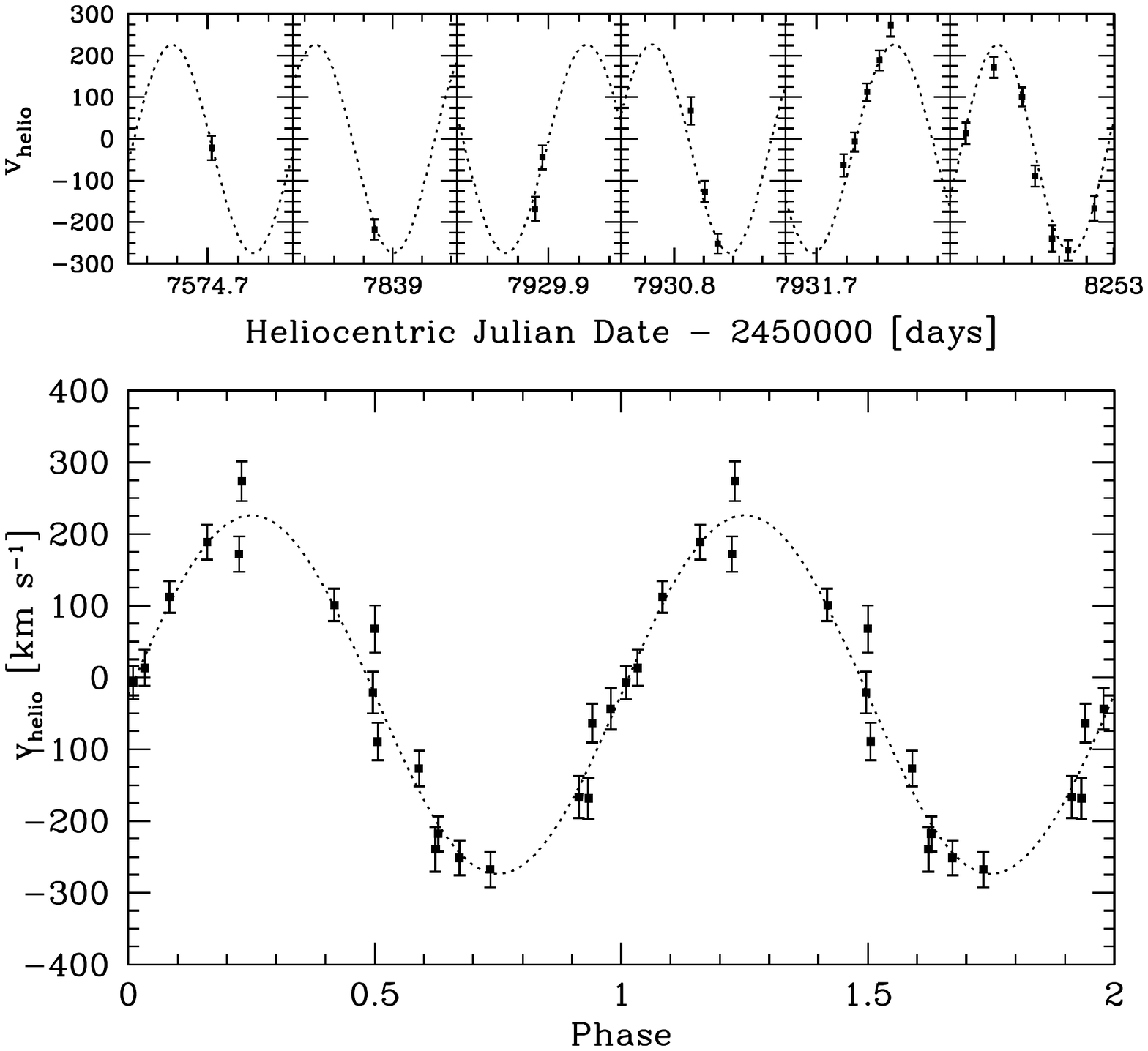}
		\caption{RV variations of SDSS\,J1618+3854, folded on the orbital period of 6.39\,hr.}
		\label{fig:j1618spec}
	\end{figure}

	We obtain our first frequency solution by adopting the highest peak in the FT near 203\,$\mu$Hz as the dominant pulsation signature.  We compute a least-squares fit of a sinusoid at this frequency to the light curve and search for additional signals in the residuals. The next highest peak is near 163\,$\mu$Hz, and we refine our fit with this sinusoid included.  Continuing in this fashion, we recognize that the next three signals correspond to harmonics and combination frequencies \citep{Brickhill1992} of the first two signals: $2f_1$, $f_1+f_2$, and $2f_1+f_2$.  These strict arithmetic relationships guide the choices between competing aliases, explaining our re-interpretation of the signal that the preliminary analysis of \citet{Bell2015} listed as an independent $f_3$. We find no additional significant signals in the residuals.  We tabulate the frequencies and amplitudes of the best-fit sinusoids in Table~\ref{tab:sol1} and overplot the solution in red in Figure~\ref{fig:data}.
	
	The dominant peak in the combined light curve from May 2014 is near 191\,$\mu$Hz, leading us to consider a second solution to the April 2014 data that selects the second-highest alias as $f_1$.  We prewhiten this and its first harmonic, which also corresponds to a peak in the FT.  The highest peak in the residuals is near 227\,$\mu$Hz. Had we not prewhitened the harmonic of $f_1$ first, we might have selected a different peak near 161\,$\mu$Hz as $f_2$, yielding a completely different frequency solution. After prewhitening the refined model, we adopt a third frequency near 149\,$\mu$Hz.  Finally, we include peaks corresponding to combination frequencies and harmonics, the precise locations of which justify the slight relaxation of our significance criterion.  This solution is given in the bottom of Table~\ref{tab:sol1} and we plot it in blue in Figure~\ref{fig:data}.

	The moral of this exercise is that one should consider exact frequency solutions from time series with gaps with skepticism. Selecting an incorrect alias can derail the prewhitening process, and tabulated solutions in the literature are usually subjective. Comparisons with asteroseismic models should try to account for this uncertainty.  Alternatively, gaps can sometimes be avoided by observing from space \citep[e.g.,][]{Hermes2017b} or from multiple sites distributed in longitude \citep[a la the Whole Earth Telescope;][]{Nather1990}. Still, we can conclude from the present data that SDSS\,J1618+3854 exhibits multiple pulsation modes with periods in the 1--2\,hr range.
	
	Encouraged by the similarity between these periods and the variability timescales of known pulsating ELM WDs \citep{Hermes2013b}, we targeted SDSS\,J1618+3854 for follow-up spectroscopy from the MMT.  The RVs measured from two spectra obtained in July 2016 and March 2017 differed by 200\,km\,s$^{-2}$.  We obtained ten additional spectra in June 2017, confirming that this is a single-lined spectroscopic binary. Seven more spectra from May 2018 improve our coverage of orbital phase, eliminating period aliases. A bootstrap analysis constrains the orbital period to $0.2664 \pm 0.0015$\,days ($6.39\pm0.04$\,h), with a semi-amplitude of $249.2\pm10.4$\,km\,s$^{-1}$. The RV measurements folded on this period is displayed in Figure~\ref{fig:j1618spec}.  The observed photometric signals are not harmonics of the candidate RV periods, so they really are pulsations rather than signatures of a photometric binary. The orbital period is too short to permit a main sequence primary star, so we conclude that SDSS\,J1618+3854 is a bona fide pulsating ELM WD.  Assuming an edge-on orientation, the minimum mass of the unseen companion is 0.67\,\msun.

	\subsection{SDSS\,J1131$-$0742}
	
	We selected SDSS\,J1131$-$0742 as an observing target based on excess scatter in its Catalina Sky Survey \citep[CSS;][]{Drake2009} photometry. Observations over two nights in January 2015 revealed dramatic peak-to-peak variations as large as 60\%, confirming this as a pulsating star.  A segment of the light curve and the FT are displayed in the second row of Figure~\ref{fig:data}. We list the frequencies of the highest significant peaks, revealed through successive prewhitening, in Table~\ref{tab:sol2}. 
	
	\begin{table}[b]
		\centering
		\caption{Frequency solutions for the January 2015 observations of SDSS\,J1131$-$0742}
		\label{tab:sol2}
		\begin{tabular}{lccr} 
			\hline
			Mode ID  & Frequency & Period & Amplitude\\
			& ($\mu$Hz) & (s) & (\%)\\
			\hline
			$f_1$ & 197.481(11) & 5063.8(3) & 23.43(5)\\
			$f_2$ & 270.33(14) & 3699(2) & 1.86(5)\\
			$2f_1$ & 394.96(3) & 2531.9(2) & 8.71(5)\\
			$3f_1$ & 592.44(7) & 1687.9(2) & 3.62(5)\\
			$4f_1$ & 789.92(17) & 1265.9(3) & 1.56(5)\\
			\hline
			
		\end{tabular}
	\end{table}
	
	\begin{figure}[t]
		\centering
		\includegraphics[height=0.85\columnwidth,angle=-90,trim={0 0 0 .5cm},clip]{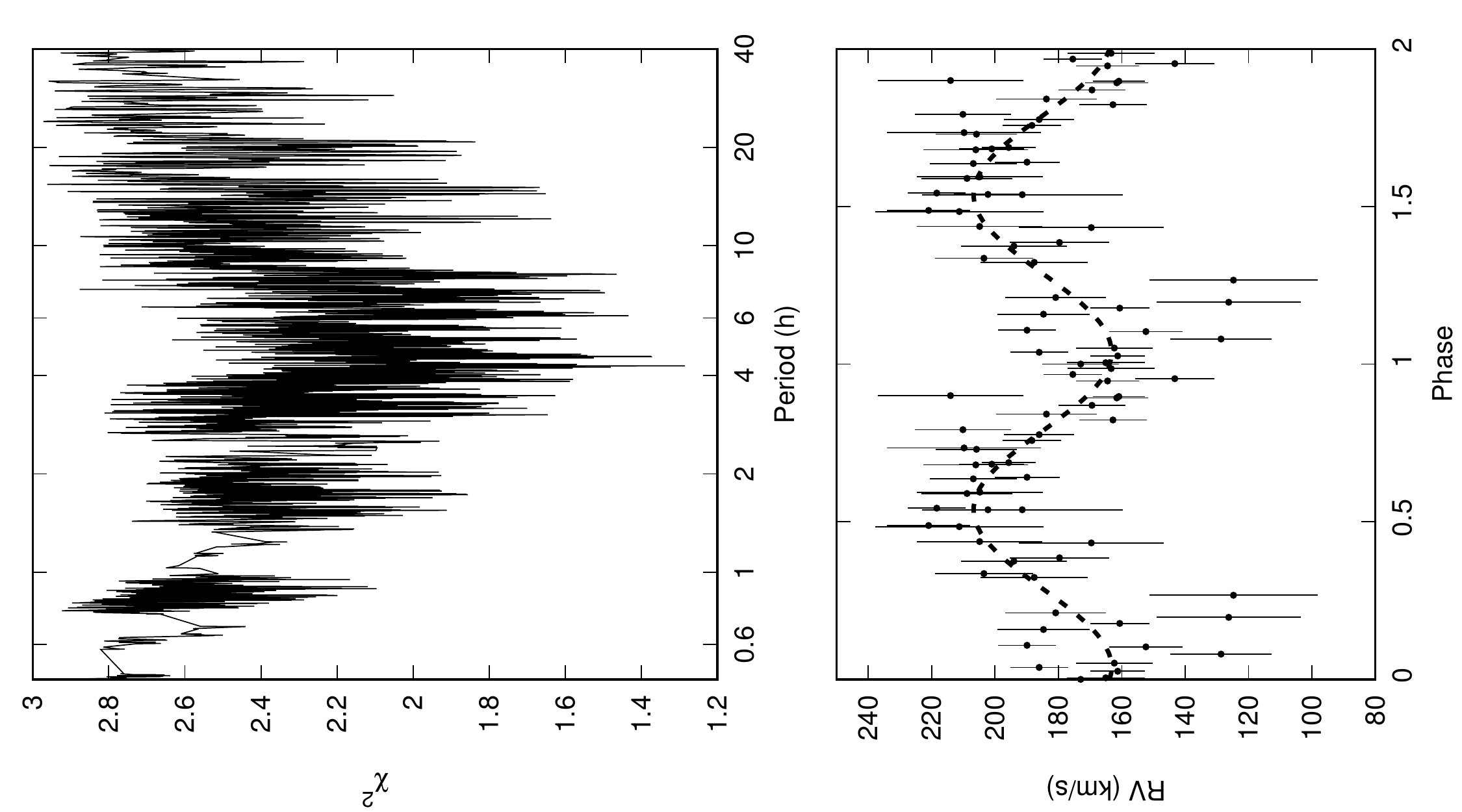}
		\caption{Radial velocity variations of SDSS\,J1131-0742. Top: $\chi^2$ of fits of different periods to the RV data.  Bottom: RV measurements folded on the best-fit period of 4.58\,hr. }
		\label{fig:j1131spec}
	\end{figure}

	\begin{figure*}[t]
		\centering
		\includegraphics[width=2\columnwidth]{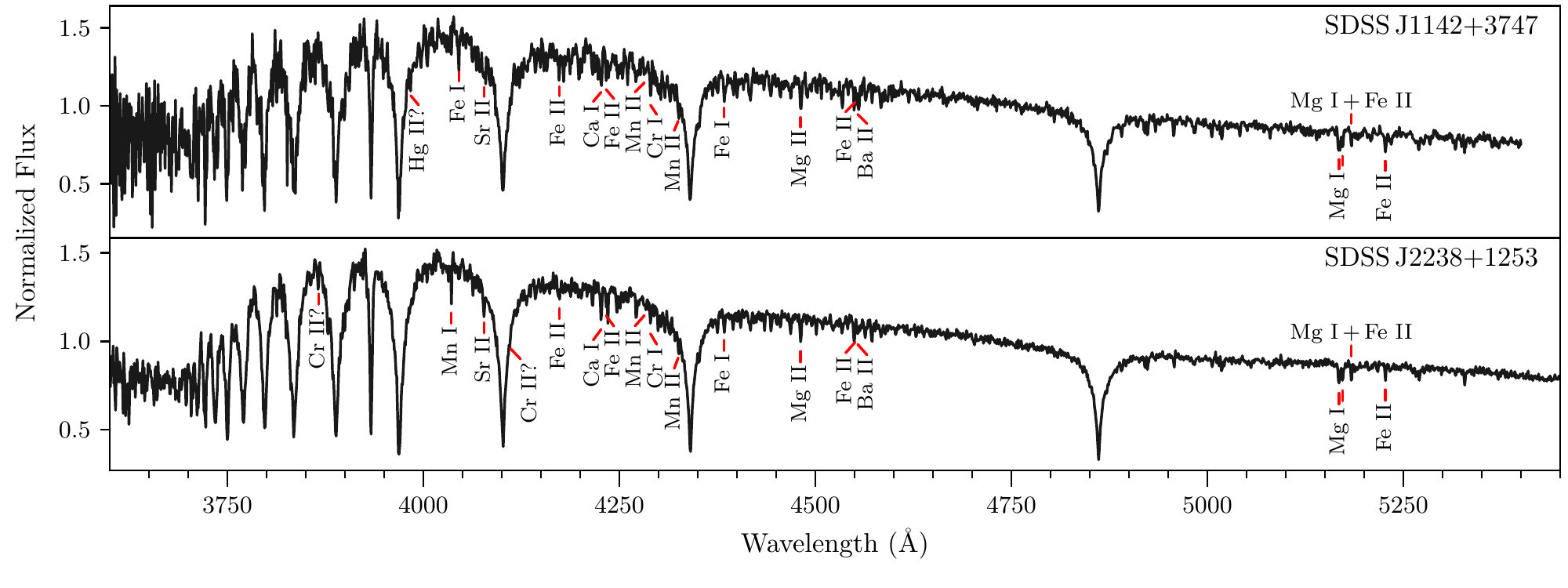}
		\caption{Summed FAST spectra of SDSS\,J1142+3747 (top) and SDSS J2238+1253 (bottom) reveal many metal lines, with some candidate line identifications indicated.}
		\label{fig:spectra}
	\end{figure*}

	SDSS\,J1131-0742 was classified as a $\delta$ Scuti by \citet{Palaversa2013} based on data from the LINEAR survey. The two dominant modes of high-amplitude $\delta$ Scuti variables (radial fundamental and first overtone) exhibit period ratios in the range 0.76--0.80 \citep[e.g.,][]{Poretti2005,Pigulski2006}. The period ratio of the two independent modes listed in Table~\ref{tab:sol2} fall just above this range; however, there is good agreement if the intrinsic frequency for $f_2$ is lower by the typical daily aliasing spacing of $\sim11.6\,\mu$Hz.

	We obtained 25 FAST spectra and 21 SOAR spectra of SDSS\,J1131$-$0742 to search for any RV signatures of binarity.  A Shapiro-Wilk test rejects the null hypothesis that the RV measurements were drawn from a Gaussian distribution with a 95\% confidence. The orbital period that minimizes $\chi^2$ is $4.28\pm0.04$\,hr (with uncertainties from Monte Carlo simulations; top panel of Figure~\ref{fig:j1131spec}). The RV semi-amplitude of $23.4\pm5.4$\,km\,s$^{-1}$ is comparable to the uncertainty on the individual measurements, and while the detection is statistically significant and phases across data from both instruments, it would be worth confirming with more precise follow-up. We plot the $\chi^2$ statistic against orbital period and the phase-folded RV data over this model in the bottom panel of Figure~\ref{fig:j1131spec}. Only compact objects can fit in such a short-period binary configuration.  The minimum mass of the unseen companion is 0.021\,\msun, and it would be 0.13\,\msun\ at an orbital inclination of 15$^\circ$.  With $\log{g}\approx4.6$ from fits of both sets of models to the SDSS spectra, we consider this to most likely be a pulsating pre-ELM WD. We note, however, that fits of the pure-H models of \citet[][and references therein]{Gianninas2011,Gianninas2014} to the summed FAST spectrum yields a higher value of $\log{g} = 5.35\pm0.13$.

	\subsection{SDSS\,J1142+3747}
	
	Our 2-hour run in April 2015 revealed marginal evidence of photometric variability in SDSS\,J1142+3747.  We revisited this object over two nights in May 2017. These data are displayed in the third row of  Figure~\ref{fig:data}.  
	
	We detect a single significant periodicity in these data, the strongest alias of which is at $263.8\pm0.2\,\mu$Hz ($3791\pm3$\,s period) with an amplitude of $0.51\pm0.04$\%. This signal timescale is consistent with pulsations observed in ELM WDs, but with only one period detected, it could also be the signature of ellipsoidal variations. ELM WDs in tight binaries often show a photometric signal at half the orbital period as the projected area of a tidally distorted star varies through the orbit \citep{Hermes2014,Bell2017}.  We obtained continuous FAST spectra over 2.1 hours---covering twice the photometric period---and found no significant RV variations to a limit of $\sim$10\,km\,s$^{-1}$.  This rules against the ellipsoidal variation interpretation, supporting that SDSS\,J1142+3747 is a pulsating star.
	
	Interestingly, the high-S/N summed FLWO spectrum of SDSS\,J1142+3747, displayed in the top panel of Figure~\ref{fig:spectra}, reveals a bounty of metal lines in addition to the deep hydrogen signatures. We mark possible identifications for a few lines by comparing against the NIST Atomic Spectra Database \citep{NIST}. By comparison with examples from the classification atlas of \citep{GrayAtlas}, SDSS\,J1142+3747 may be a chemically peculiar Am or Ap star \citep[e.g.,][]{Preston1974}, some of which exhibit $\delta$ Scuti pulsations with a similar timescale \citep[e.g.,][]{Balona2011a,Balona2011b}. We prefer this main-sequence explanation for SDSS\,J1142+3747, as the spectroscopic models used to fit the SDSS data do not account for peculiar atmospheres, and the inferred \logg\ could be inaccurate as a result.
	
	\subsection{SDSS\,J2238+1253}
	
	Our analysis and interpretation of SDSS\,J2238+1253 follows that of the previous object, SDSS\,J1142+3747, very closely.  This target also exhibits a single independent photometric signal with a similar period.  The two peaks marked in the FT in the fourth row of Figure~\ref{fig:data} are the highest alias and its second harmonic: $f_1 = 267.114\pm0.005\,\mu$Hz ($3743.72\pm0.07$\,s periodicity) with an amplitude of $4.329\pm0.018$\% (harmonic amplitude at $0.488\pm0.018$\%).
	
	As with the previous object, we considered that a single dominant period could correspond to ellipsoidal variations, though the observed amplitude in this case is larger even than for the shortest-period (12.75\,min) detached ELM WD binary known \citep{Brown2011}. We obtained 2.1-hours of continuous FAST spectra that do not reveal any RV variability to a limit of $\sim$10\,km\,s$^{-1}$. SDSS\,J2238+1253 was targeted by the parallel effort of \citet{Pelisoli2018b} to characterize sdAs with time series spectra, who also found no evidence of RV variability in the period range of 20\,min to 180\,days. 
	
	The summed FAST spectrum of SDSS\,J2238+1253 displayed in the bottom panel of Figure~\ref{fig:spectra} also shows many metal lines, which might be causing a blanketing effect that is unaccounted for in our atmospheric models. This may also be a chemically peculiar star \citep{GrayAtlas}. We draw the same conclusion as for SDSS\,J1142+3747 (top panel of Figure~\ref{fig:spectra}), that SDSS\,J2238+1253 is likely a $\delta$ Scuti pulsator on the main sequence with an overestimated \logg\ caused by missing physics in the applied models.

	\subsection{SDSS\,J1310$-$0142}
	
	Our light curve of SDSS\,J1310$-$0142 from 14 Mar 2015 showed some evidence of photometric variability.  We confirmed that this target is a multi-periodic pulsator over three consecutive nights in June 2017.  The FT and a portion of the light curve are displayed in the fifth row of Figure~\ref{fig:data}, and the two independent signals, plus a harmonic of the dominant mode, are detailed in Table~\ref{tab:j1310sol}.  The FT shows broad aliasing structure, but the most likely signals do not yield a period ratio in the typical range for $\delta$ Scuti variables \citep[0.76--0.80; e.g.,][]{Poretti2005,Pigulski2006}. \citet{SA2018} compare pulsation periods calculated for theoretical $\delta$ Scuti and pre-ELM models to develop asteroseismic tools to distinguish these classes, but these are difficult to apply to the current observations of SDSS\,J1310$-$0142 given the considerable aliasing in the spectral window.  Additional future observations would be worthwhile to improve constraints on the pulsation periods and the models that can explain them.

	\begin{table}
		\centering
		\caption{Frequency solutions for the June 2017 observations of SDSS\,J1310$-$0142}
		\label{tab:j1310sol}
		\begin{tabular}{lccr} 
			\hline
			Mode ID  & Frequency & Period & Amplitude\\
			& ($\mu$Hz) & (s) & (\%)\\
			\hline
			$f_1$ & 318.25(15) & 3142.2(14) & 1.21(6)\\
			$f_2$ & 481.7(2) & 2075.8(9) & 0.86(6)\\
			$2f_1$ & 636.5(5) & 1571.1(12) & 0.31(6)\\
			\hline
			
		\end{tabular}
	\end{table}

	\subsection{SDSS\,J1604+0627}
	
	The photometric variability observed in SDSS\,J1604+0627 is quite different in character from the other new variable sdAs.  The frequency of the single significant signal that we detect in this star is much higher, at $3651.35\pm0.19\,\mu$Hz ($273.871\pm0.014$\,s period) with amplitude $0.37\pm0.02$\%.  To help make the pulsations visible to the eye, we plot a running average over 10 adjacent measurements in the sixth row of Figure~\ref{fig:data}.  The displayed FT for this object runs to higher frequency in order to contain the single significant periodicity that we detect.
	
	There are only a couple of known types of pulsating star that show pulsation periods this short.  The observed Balmer lines are too narrow for this to be a low-order gravity-mode (g-mode) in a pulsating white dwarf of typical mass.  Rapidly Oscillating Ap stars (roAp) exhibit short, high-overtone pressure-modes (p-modes), but a periodicity as short as $274$\,s has never been observed \citep{Kurtz2006}. P-modes in hot subdwarfs could have such short periodicities; while \citet{Pelisoli2018a} suggested that some sdAs may be binaries that contain a hot sdB star, these systems must have a UV flux excess, which we rule out for SDSS\,J1604+0627 given the low (nuv$-g$) color of $\approx$2.79 from GALEX data.
	
	\citet{SA2018} compare models of pre-ELM WDs that are evolving across the main sequence ($\log{g}\approx 4$) to $\delta$ Scuti models and claim that periods below 700\,s are exclusive to pre-ELM WDs.  The pre-ELM models of  \citet{Istrate2016} also predicted p-mode pulsations with periods that agree with the 4.6-minute signal of SDSS\,J1604+0627. With spectroscopic $\log{g}\approx 5.7$ from fits of pure-H model atmospheres to the SDSS spectra, SDSS\,J1604+0627 is situated in the transition region between pre-ELM and ELM WDs in evolutionary models, and also where pre-ELM WDs may evolve repeatedly through during violent CNO-flash episodes \citep[e.g.,][]{Althaus2013}. While observed pulsation periods as short as 108\,s and 134\,s have been suggested to be the signatures of p-modes in the pulsating ELM WD SDSS\,J1112+1117 \citep{Hermes2013b}, the single period observed in SDSS\,J1604+0627 is longer than the p-mode range from models of ELM WDs \citep{Corsico2016a}. This could instead correspond to a low-$k$ g-mode that is driven by the $\epsilon$-mechanism \citep{Corsico2014b}. 
	
	Following \citet{Hermes2017}, we inspected the individual SDSS subspectra and find no evidence of RV variations on timescales of 30 minutes to 5 days, supporting that SDSS\,J1604+0627 is not part of a short-period binary system. Most ELM WDs in double-degenerate binaries are expected to merge within a Hubble time \citep{Brown2016b}, and a post-merger ELM WD was recently identified by \citet{Vos2018}. We consider SDSS\,J1604+0627 to be a good candidate for the first pulsating post-merger (pre-)ELM WD.
	
	We note, however, that a single periodicity is difficult to confirm as being caused by pulsations. For example, in reconsidering the photometric variations in hot DQ (carbon atmosphere) WDs that were originally thought to be caused by stellar pulsations \citep[begining with SDSS J1426+5752;][]{Montgomery2008}, \citet{Lawrie2013} and \citet{Williams2016} have suggested that these signals can be explained more simply by fast rotational modulation. Detection of rapid rotation in SDSS\,J1604+0627 would also be extremely interesting in the context of a post-merger formation.  Continued monitoring of SDSS\,J1604+0627 will help to reveal the nature of this signal, as we may be able to detect the secular rate of period change from pulsations, as explored in (pre-)ELM WD evolutionary models by \citet{Calcaferro2017a}.  Multi-wavelength observations could also discern between these explanations, as the characteristic colors of surface inhomogeneities from spots and stellar pulsations can be very different \citep[discussed for WDs by, e.g.,][]{Dupuis2000,Robinson1982}.
	
	The unique periodicity of this sdA star will motivate continued theoretical efforts to understand the diversity of stellar pulsations and evolutionary pathways in the parameter space of sdAs. Such a short period pulsation may be sensitive to previously unprobed regions of the interior of a star in a potentially quite rare phase of evolution toward a final ELM WD cooling track.
	
	\subsection{SDSS\,J0756+5027}
	
	SDSS\,J0756+5027 was selected as an observing target based on significant excess scatter in the CSS data \citep{Drake2009}.  Our single light curve from 11 Feb 2015 in the bottom left panel of Figure~\ref{fig:data} shows changes of more than a factor of two in brightness.  Even with more than seven hours of coverage, we do not observe a complete period of the dominant pulsation mode.

	\begin{figure*}[t]
		\centering
		\includegraphics[width=1.95\columnwidth,trim={0 0 .65cm 0},clip]{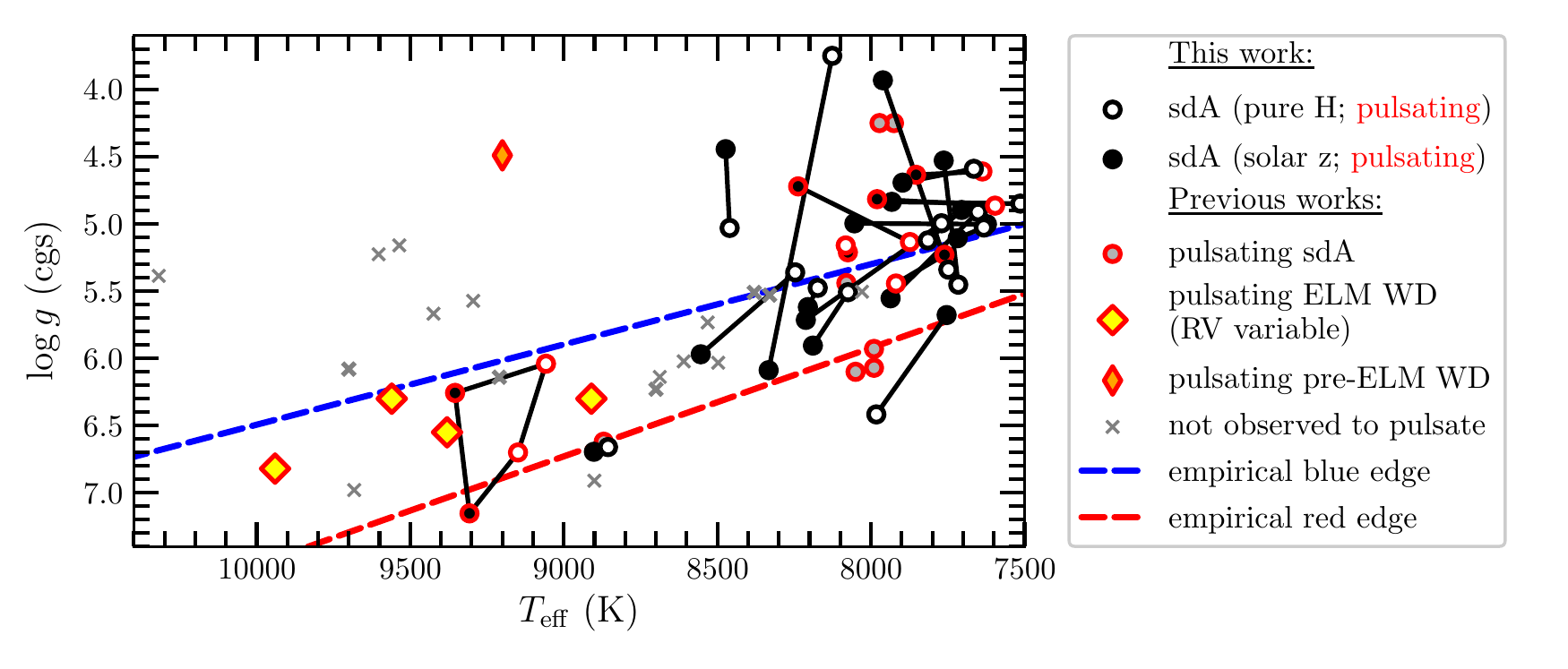}
		\caption{Spectroscopic parameters from pure-H (white-filled circle) and solar-z (black-filled) model fits to SDSS spectra are connected by lines for 23 sdA stars that we observed photometrically. We fit two spectra of SDSS\,J1618+3854, represented by four connected points in this figure. New pulsating variables are outlined in red. We indicate other variables and non-variables from previous searches for pulsating (pre-)ELM WDs for context, along with the empirical low-mass extension of the ZZ Ceti instability strip boundaries (see text).}
		\label{fig:pspace}
	\end{figure*}

	Constructing a periodogram of the CSS photometry reveals a strong periodicity of roughly 12.263 hours, which we confirm with data from ATLAS \citep{Heinze2018}. The CSS data, folded on twice this period, is displayed in the bottom right panel of Figure~\ref{fig:data}.  Folded this way, the data suggest a change in amplitude between alternate pulsations, in line with the period doubling behavior observed in RR Lyrae stars and suspected of being connected to the Blazhko effect \citep{Szabo2010}. \citet{Drake2013} classified this star as RRab-type in the catalog of RR Lyrae from CSS data. Given the high \logg\ inferred from pure-H model atmospheres, SDSS\,J0756+5027 could be physically similar to the 0.26\,\msun\ star OGLE-BLG-RRLYR-02792 that exhibits RR-Lyrae-like pulsations \citep{Piet2012}. That star is the recent product of mass transfer in a system that will eventually evolve to become a double-degenerate binary. \citet{Karczmarek2017} address the theoretical properties and context of such binary mass-transfer products. We note that the solar-z model fit yields a significantly lower $\log{g}\approxeq 4.05$ that approaches the typical range for RR Lyrae variables on the horizontal branch \citep[$\log{g}\lesssim 3.7$; e.g.,][]{Nemec2013}.

	\section{The Pulsational View of sdA Stars}
	\label{sec:disc}
	
	Out of 23 sdAs observed, we find signatures of pulsations in seven.  Their observational properties are summarized in Table~\ref{tab:sum}.  We plot the locations of all of the sdA stars that we observed in spectroscopic \logg--\teff\ space, as determined using both solar-z and pure-H models, in Figure~\ref{fig:pspace}.  The variables are outlined in red. We include a number of other stars for context: pulsating ELM WDs known to be in short-period binaries \citep{Hermes2012,Hermes2013a,Kilic2015}; a pulsating pre-ELM WD in a short-period orbit \citep{Maxted2014}; pulsating sdA stars \citep[not known to be in short period binaries;][]{Hermes2013b,Corti2016,Bell2017}; and stars from the ELM Survey with limits on a lack of pulsations \citep[<1\% variability; some RV-variable and some not;][]{Hermes2012,Hermes2013a,Hermes2013b,Bell2017}. The empirical boundaries of the H-atmosphere WD instability strip from \citet{Gianninas2015} are marked with dashed lines. Corrections have been applied to the spectroscopic parameters of all stars with $\log{g}\geq5.0$ based on 3D hydrodynamic simulations of convection in WD atmospheres \citep{Tremblay2013,Tremblay2015}.
	
	The pulsation timescales and orbital constraints from time series spectroscopy separate our seven new pulsating stars into several categories:
	\begin{itemize}
		\item SDSS\,J1618+3854 is a binary system with a 6.39-hour orbital period that can only accommodate compact objects, and we also make a marginal detection of 4.58-hour RV variations from SDSS\,J1131$-$0742. The pulsation periods for both are in the 1--2 hour range, and the peak-to-peak variations are as high as 60\% for SDSS\,J1131$-$0742. Given their locations in spectroscopic parameter space, the latter object may be a pre-ELM WD, while we suspect that SDSS\,J1618+3854 is on its final cooling track given its high spectroscopic \logg\ compared to evolutionary models \citep[e.g.,][]{Althaus2013,Istrate2016a}.
		
		\item SDSS\,J1142+3747 and SDSS\,J2238+1253 each exhibit a single independent pulsation signal of $\sim$1\,hr. We do not measure significant RV variations in their spectroscopic data, but we do detect many metal lines. We interpret these targets as possibly chemically peculiar main-sequence stars that exhibit $\delta$ Scuti pulsations.
		
		\item The FT of SDSS\,J1310$-$0142 reveals two independent pulsation periods, which may be used to discern between $\delta$ Scuti and pre-ELM WD pulsations \citep{SA2018} if additional observations can achieve less aliasing in the spectral window.
		
		\item SDSS\,J1604+0627 has a photometric period of 4.56\,min that is unique in the sdA regime. With additional theoretical investigation, this signal may help to resolve the evolutionary state of SDSS\,J1604+0627, which lies in a \logg--\teff\ region of overlap of stellar models that are transitioning between contracting pre-ELM WDs and cooling-track ELM WDs, and pre-ELM WDs that are executing extremely rapid loops during episodic CNO flashes \citep[e.g.,][]{Althaus2013,Istrate2016a}.  SDSS subspectra indicate that this star is not in a short-period binary, so it may be a post-merger object.
		
		\item The dominant pulsation period of SDSS\,J0756+5027 is 12.3\,hr, and we interpret from the light curve morphology that this star is undergoing RR-Lyrae-type pulsations. It may be a pre-ELM WD like the low-mass RR-Lyrae-type variable OGLE-BLG-RRLYR-02792 \citep{Piet2012}.
	\end{itemize} 
	The diversity in pulsation properties of the variable sdA stars supports the emerging narrative that this population comes from a mixture of formation and evolution scenarios.  It also reveals that it is an asteroseismically rich population, with many of the sub-classes exhibiting pulsations.  More detailed follow-up analyses of the observed pulsational eigenfrequencies will enable us to constrain the interior structures of these stars, including the products of mass transfer in tight binaries.
	
	SDSS\,J1618+3854 is an important addition to the pulsating ELM WD class that has been growing steadily since the first discovery by \citet{Hermes2012}. Based on the preliminary characterization of SDSS\,J1618+3854 presented by \citep{Bell2015}, \citet{Calcaferro2017b} found that the pulsation periods of SDSS\,J1618+3854 agree best with calculations for an ELM WD model with $M_\star=0.1706$\,\msun\ and $T_\mathrm{eff} = 9076$\,K. \citet{Bell2015} suggested that the nonlinearities in the pulsations of SDSS\,J1618+3854 make it a good candidate for asteroseismically constraining the behavior of stellar convection in the new physical regime of ELM WDs, following the method of \citet{Montgomery2005}.  We maintain that this would be a valuable analysis, but it requires that we pin down the intrinsic pulsation frequencies from amongst the aliases, which likely will require a multi-site observing campaign.
	
	\begin{table*}
		\caption{Summary of new pulsating sdA star properties}              
		\label{tab:sum}      
		\centering                                      
		\begin{tabular}{l l l l l l}  
			\hline\hline
			Star & \teff & \logg & pulsation period & RV variable? & suggested class\\
			(SDSS) & (K; pure H) & (cgs; pure H) & range (s) \\
			\hline
			
			J1618+3854\tablefootmark{a} & 9149 &6.70 & 5000--6100 & yes & ELM WD\\
			J1131$-$0742 & 7637&  4.61 & 3700--5100 & yes (?) & pre-ELM (?)\\
			J1142+3747 & 8082&  5.16 & 3800 & no & $\delta$ Scuti (?)\\
			J2238+1253 & 7999&  5.03 & 3700 & no & $\delta$ Scuti (?)\\
			J1310$-$0142 & 8224&  5.33& 2100--3100 & ? & pre-ELM/$\delta$ Scuti (?)\\
			J1604+0627 & 8097&  5.71 & 274 & no & pre-ELM/ELM WD (?)\\
			J0756+5027 & 6875 & 5.34 & 44100 & ? & ELM WD/RR Lyrae (?)\\
			\hline
			
		\end{tabular}
		\tablefoot{
			\tablefoottext{a}{See Table~\ref{tab:targets} for second set of spectroscopic parameters.}}
	\end{table*}

	Pre-ELMs represent an exciting phase of stellar evolution, where many stars might execute extremely rapid CNO-flash loops that burn residual hydrogen \citep[those with masses $0.18\,M_{\odot}\lesssim M_\mathrm{WD} \lesssim 0.4\,M_{\odot}$ in the models of][]{Althaus2013}. The ELM WD precursors can be quite large before they reach the cooling track; e.g., \citet{Piet2012} constrain the radius of the low-mass RR-Lyrae-type variable OGLE-BLG-RRLYR-02792 to be $>4$\,\rsun\ from eclipses. Three of our new variables exhibit exotic variations that may provide particular insight into pulsational driving and interior structures in the pre-ELM WD regime.  The high-amplitude pulsations of SDSS\,J1131$-$0742, the 4.56-min period of SDSS\,J1604+0627, and the RR-Lyrae-type variability of SDSS\,J0756+5027 all warrant additional theoretical investigations to sufficiently explain. Existing theoretical diagnostics for asteroseismically discerning between pre-ELM WDs and $\delta$ Scuti variables from \citet{SA2018} insist that the short period observed in SDSS\,J1604+0627 cannot belong to a $\delta$ Scuti star.
	
	Preliminary constraints on sdA radii from \emph{Gaia} DR2 parallaxes indicate that hundreds are ELM WDs or their precursors \citep{Pelisoli2018c}. Additional analyses and data from future \emph{Gaia} releases will further disentangle the properties of the multiple sdA subpopulations. Independent radius measurements for pulsating sdAs and ELM WDs will also inform our asteroseismic constraints on stellar structures, which is especially important for stars that exhibit only a few pulsation modes, such as those presented here.
	
	In parallel and in agreement with the present work, \citet{Pelisoli2018b} detected pulsations with $\sim$1--2\,hr periods in time series photometry of seven out of 21 additional sdA stars that they observed.
	
	To enable more detailed asteroseismic analyses, we are publishing the reduced light curves of all new variable sdAs as a supplement to this manuscript.

	\begin{acknowledgements}
		
		We thank the referee for helpful comments that improved this manuscript. We thank Alex Gianninas for fitting models to the FAST spectrum of SDSS\,J1131$-$0742, Michel Breger for many helpful discussions about $\delta$ Scuti variables, and Arumalla B.~S.~Reddy for assistance in identifying metal lines in FAST spectra.  We also thank Alejandro C\'{o}rsico, Leila Calcaferro and Leandro Althaus for inspecting the pulsation properties of their CNO-flashing evolutionary models.
		K.J.B., D.E.W., K.I.W.,  Z.V., B.G.C. and M.H.M.\ acknowledge support from NSF grant AST-1312983. 
		K.J.B.\ was also supported by the European Research Council under the European Community's Seventh Framework Programme (FP7/2007-2013) / ERC grant agreement no 338251 (StellarAges) during the preparation of this manuscript. I.P.\ acknowledges support from CNPQ-Brazil. Support for this work was provided by NASA through Hubble Fellowship grant \#HST-HF2-51357.001-A, awarded by the Space Telescope Science Institute, which is operated by the Association of Universities for Research in Astronomy, Incorporated, under NASA contract NAS5-26555. A preliminary version of this work was included in K.J.B.'s PhD thesis (U.\ Texas).
		This paper includes data taken at The McDonald Observatory of The University of Texas at Austin. Thank you to the McDonald observing support, esp.\ David Doss, Coyne Gibson, and John Kuehne. The authors acknowledge the Texas Advanced Computing Center (TACC) at The University of Texas at Austin for providing database resources that have contributed to the research results reported within this paper. 
		The CSS survey is funded by the National Aeronautics and Space
		Administration under Grant No. NNG05GF22G issued through the Science
		Mission Directorate Near-Earth Objects Observations Program.  The CRTS
		survey is supported by the U.S.~National Science Foundation under
		grants AST-0909182 and AST-1313422.
		
	\end{acknowledgements}


\end{document}